\documentclass[]{spie}  

\usepackage{amsmath,amsfonts,amssymb}
\usepackage{graphicx}
\usepackage[colorlinks=true, allcolors=blue]{hyperref}

\title{Estimation of polarization aberrations and their effect on the coronagraphic performance for future space telescopes}

\author[a]{Ramya M Anche}
\author[a]{Sebastiaan Y. Haffert}
\author[a,b]{Jaren N Ashcraft}
\author[a,b]{Kian Milani}
\author[a]{Kyle Van Gorkom}
\author[a,b]{Kevin Derby}
\author[a]{Ewan S. Douglas}
\author[c]{Maxwell A. Millar-Blanchaer}
\affil[a]{Steward Observatory, University of Arizona, 933N Cherry Avenue, Tucson, Arizona, 85721, USA}
\affil[b]{James C. Wyant College of Optical Sciences, University of Arizona, 933N Cherry Avenue, Tucson, Arizona, 85721, USA}
\affil[c]{Department of Physics, University of California, Santa Barbara, CA, 93106, USA}

\authorinfo{Further author information: (Send correspondence to R.M.A.)\\R.M.A.: E-mail: ramyaanche@arizona.edu}

\begin{document} 
\maketitle

\begin{abstract}
     A major goal of proposed future space observatories, such as the Habitable World Observatory, is to directly image and characterize Earth-like planets around Sun-like stars to search for habitability signatures requiring the starlight suppression (contrast) of $10^{-10}$. One of the significant aspects affecting this contrast is the polarization aberrations generated from the reflection from mirror surfaces. The polarization aberrations are the phase-dependent amplitude and phase patterns originating from the Fresnel reflections of the mirror surfaces. These aberrations depend on the angle of incidence and coating parameters of the surface. This paper simulates the polarization aberrations for an on-axis and off-axis TMA telescope of a 6.5 m monolithic primary mirror. We analyze the polarization aberrations and their effect on the coronagraphic performance for eight different recipes of mirror coatings for Astronomical filter bands g-I: three single-layer metal coatings and five recipes of protective coatings. First, the Jones pupils are estimated for each coating and filter band using the polarization ray tracing in Zemax. Then, we propagate these Jones pupils through a Vector Vortex Coronagraph and Perfect Coronagraphs using hcipy, a physical optics-based simulation framework. The analysis shows that the two main polarization aberrations generated from the four mirrors are the retardance-defocus and retardance-tilt. The simulations also show that the coating plays a significant role in determining the strength of the aberrations. The bare/oxi-aluminum and Al+18nm LiF coating outperforms all the other coatings by one order of magnitude. 
\end{abstract}


\section{INTRODUCTION}
\label{sec:intro}  
In any optical system, when the light gets reflected/transmitted from the optical element, the differences in the Fresnel reflection/transmission coefficients generate polarization aberrations. These polarization aberrations depend on the curvature and the coating of the optical element \cite{mcguire1987polarization,chipman1987polarization,chipman1989polarization,mcguire1994polarization,mcguire1994polarization2}. In the field of astronomy, instrumental polarization, and crosstalk arising from these aberrations can be calibrated/mitigated while performing polarimetry \cite{sanchez1992instrumental}; however, they are difficult to compensate for in the high-contrast imaging \cite{breckinridge2015polarization,van2021high,mendillo2019polarization,breckinridge2018terrestrial}. The presence of the aberrations, retardance tilt and defocus has been observed in the high contrast imaging data from the ground-based instruments, Gemini/GPI and VLT/SPHERE/ZIMPOL \cite{millar2022polarization,schmid2018sphere}.

As recommended by the Astro2020 Decadal survey, the proposed Habitable World Observatory aims to reach a raw contrast of $10^{-10}$ in optical wavelengths to search for the biomarkers and characterize earth-like exoplanets with a 6.5m telescope and a coronagraph/starshade. Polarization aberrations generated from the mirror surfaces could be one of the limiting factors for achieving contrast, as shown previously for optical designs of LUVOIR and Habex \cite{breckinridge2018terrestrial,davis2018habex,will2019effects,davis2019polarization}. Thus, modeling polarization aberrations and estimating their effect on the contrast must be performed before designing coronagraph optics to develop a mitigation strategy.

The Modeling of polarization aberrations for the next-generation GSMTs by \textit{Anche et al. (2023)} \cite{anche2023polarization,anche2018estimation} revealed that coating plays a significant role in the magnitude of these aberrations; The bare aluminum coating performs much better for the Giant Magellan Telescope than the Gemini-like coating, although the latter is optimized for higher reflectivity. Our simulations show that the peak raw contrast for GMT with Bare aluminum coating is in the order of $10^{-4}$ and becomes an order worse $\sim$ $10^{-3}$ for Silver and Gemini coating for a fourth-order perfect coronagraph.

In this paper, we present the effect of polarization aberrations on the achievable contrast for ten different coating recipes for two telescopes designed as a concept study for next-generation space telescopes. The two telescopes considered are three-mirror anastigmats with a fourth mirror used to fold the beam. We also study the polarization aberration residuals at the focal plane of different coronagraph architectures: Perfect coronagraph of order 2, 4, and 6 \cite{anche2023polarization}, Vector Vortex coronagraph \cite{mawet2009vector}, and Scalar Vortex Coronagraph \cite{ruane2019scalar}  of charge 2,4 and 6. We aim to understand the combined effect of coating and coronagraphs to determine the best coating+coronagraph combination for future space telescopes. Section \ref{sec1} details the telescope's optical design and different coating recipes used. The Jones pupil maps generated using the polarization ray tracing algorithm are in Section \ref{sec2}. The description of the coronagraphic masks and the results on the effect of the contrast is shown in Section \ref{sec3}. Finally, the summary and future work are in Section \ref{sec4}.
\section{Modelling of polarization aberrations}
\label{sec1}
We model the polarization aberrations using the polarization ray tracing (PRT) algorithm, where the rays are traced from the primary mirror to the focal plane, calculating the change in polarization at every optical surface \cite{chipman1989polarization,chipman1995mechanics}. The ray tracing of the optical system is performed using Ansys Zemax Optics studio, and the estimation of reflection coefficients and Jones pupils for different coatings is performed in Python and Poke \cite{Ashcraft_poke_2022,Ashcraft_poke_2023}. Poke is an open-source Python package that uses the Zemax OpticStudio api to trace rays and perform PRT in a Python environment. The full description of the PRT algorithm used in our simulations is explained in detail in \cite{chipman2018polarized} and \cite{anche2023polarization}.
\subsection{PRT: Optical layout, incident angles, reflection coefficients}
\begin{figure}[!h]
\centering
\includegraphics[width=0.98\textwidth]{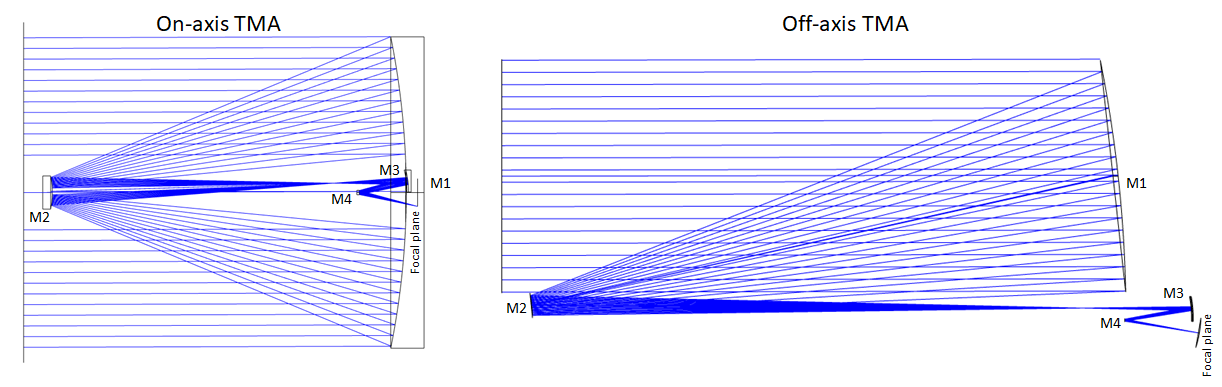}
\caption{Optical layout of the telescopes from Zemax\textsuperscript{\textregistered} used in the simulations of polarization aberrations}
\label{opl}
\end{figure}

We consider the two telescope designs for our simulations: an on-axis three-mirror anastigmat and an off-axis three-mirror anastigmat, as shown in Figure \ref{opl}. Each telescope design is a conventional three-mirror TMA design with an ellipsoidal primary mirror, a hyperbolic secondary mirror, and an ellipsoidal tertiary. The fourth mirror is a fold mirror used to fold the beam. The primary mirror in both these telescopes is a monolithic mirror of 6.5m diameter. The on-axis TMA has a  primary of F/1.27, with 8m total optical length and a final beam of F/15, whereas the off-axis TMA design is based on LUVOIR-B design with a final beam of F/35 has a primary mirror of F/2.81.
The optical parameters of the on-axis TMA are provided in \textit{Daewook et al. (2023)}\cite{daewook2023, heejoo2023}, and the off-axis TMA is provided in \textit{Nicholas et al. (2023)}\cite{Nicholas2023}. Both these telescopes are optimized to provide diffraction-limited performance in the optical wavelength regions. The higher-order optical aberrations are present, which are expected to be compensated using wavefront sensing and control.  
\par
\begin{figure}[!h]
\centering
\includegraphics[width=0.45\textwidth]{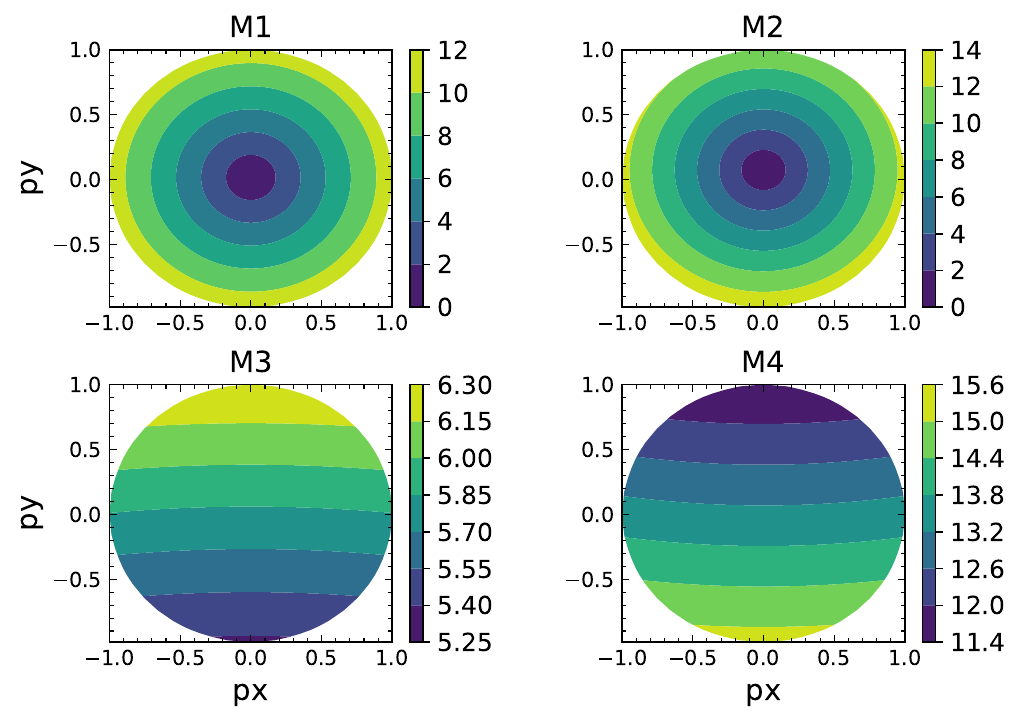}
\includegraphics[width=0.45\textwidth]{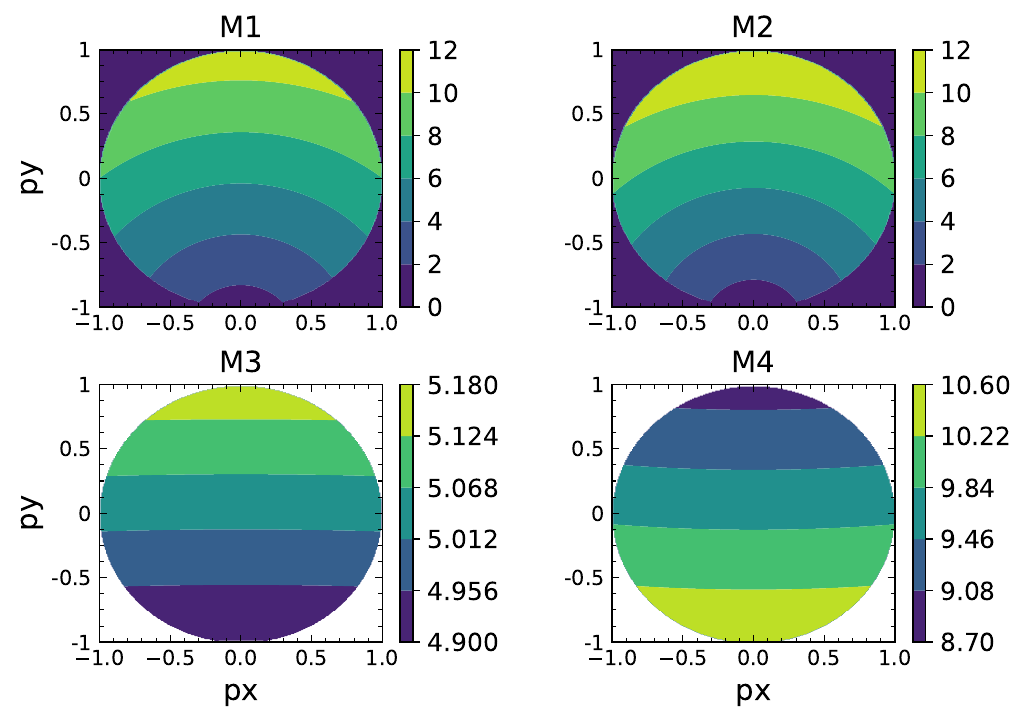}
\caption{The incident angles for all four mirrors for the on-axis TMA (left-panel) and the off-axis TMA (right-panel) are shown. The x and y axis are the normalized pupil coordinates.}
\label{inc-ang}
\end{figure}

\begin{figure}[!h]
    \centering
    \includegraphics[width=0.8\textwidth]{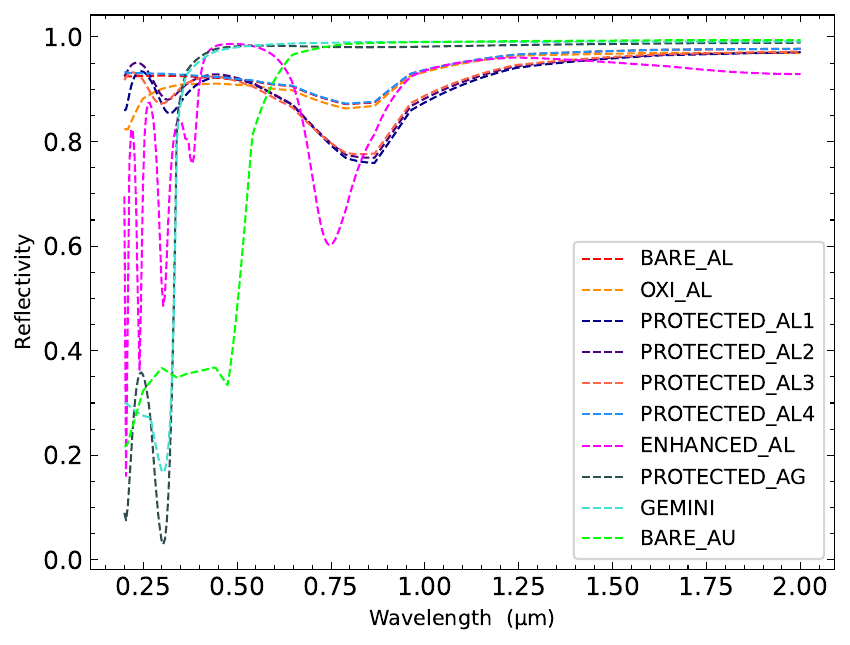}
    \caption{Reflectivity estimated for different coating recipes is shown over the wavelength. aluminum and Protected aluminum recipes are seen to show high reflectivity in the entire wavelength region.}
    \label{ref-all-coat}
\end{figure}

 Using the raytracing in Zemax, we obtain the AOI on all the mirror surfaces, as shown in Figure \ref{inc-ang}. For the M1 and M2 of the On-axis TMA, the AOI varies from 0 to 14\textdegree~from the center to the edge of the mirror, whereas the AOI on M3 and M4 vary linearly from one end to the other end. For off-axis TMA, the AOI of M1 and M2 varies from 0 to 12\textdegree~from one end to the other end of the mirror due to the off-axis profile, while for M3 and M4, the AOI varies from 0.28\textdegree and 1.9\textdegree~from one end to the other end. These telescopes have small AOI variations and avoid 45\textdegree~fold mirror following the design rules defined by \textit{Breckinridge et al. (2015)} for minimizing polarization aberrations.  
The Fresnel reflection coefficients are estimated using the thin-film algorithm from \textit{Macleod (2010)}\cite{macleod2010thin} for ten different coating recipes. We use the same coating recipe for two telescopes on all four mirror surfaces. The refractive indices used in our simulation are provided in the Appendix. The reflectivity estimated for all the coating recipes is shown in Figure \ref{ref-all-coat}. Bare aluminum and Protected aluminum recipes have reflectivity $>$ 80\% for the wavelength ranging from 0.2-2$\mu$ m. The Enhanced aluminum coating is optimized to have $>$ 95\% reflectivity in the optical wavelengths; hence, the reflectivity falls to 60\% in the near-IR region. Protected Silver and Gemini show $>$95\% reflectivity from 0.3-2 $\mu$m but $<$ 50\% below 0.3$\mu$m. Bare gold performs well beyond 0.55$\mu$m but shows $<$ 50\% reflectivity in the UV region. 
\begin{itemize}
    \item Bare aluminum and Bare Gold
    \item Oxidized aluminum: aluminum+5nm $\rm Al_2O_3$ 
    \item Protected\textunderscore AL1: aluminum+174nm $\rm SiO_2$ 
    \item Protected\textunderscore AL2: aluminum+174nm $\rm MgF_2$
    \item Protected\textunderscore AL3: aluminum+174nm $\rm LiF$
    \item Protected\textunderscore AL4: aluminum+18nm $\rm LiF$
    \item Enhanced\textunderscore AL: aluminum+ 54.7 nm $\rm TiO_2$+ 87 nm $\rm SiO_2$+54.7 nm $\rm TiO_2$+ 87 nm $\rm SiO_2$
    \item Protected\textunderscore AG: Silver+150nm $\rm SiO_2$+30nm $\rm Al_2O_3$
    \item Gemini: Silver+8.5nm $\rm Si_3N_4$
\end{itemize}

\begin{figure}[!h]
    \centering
    \includegraphics[width=0.95\textwidth]{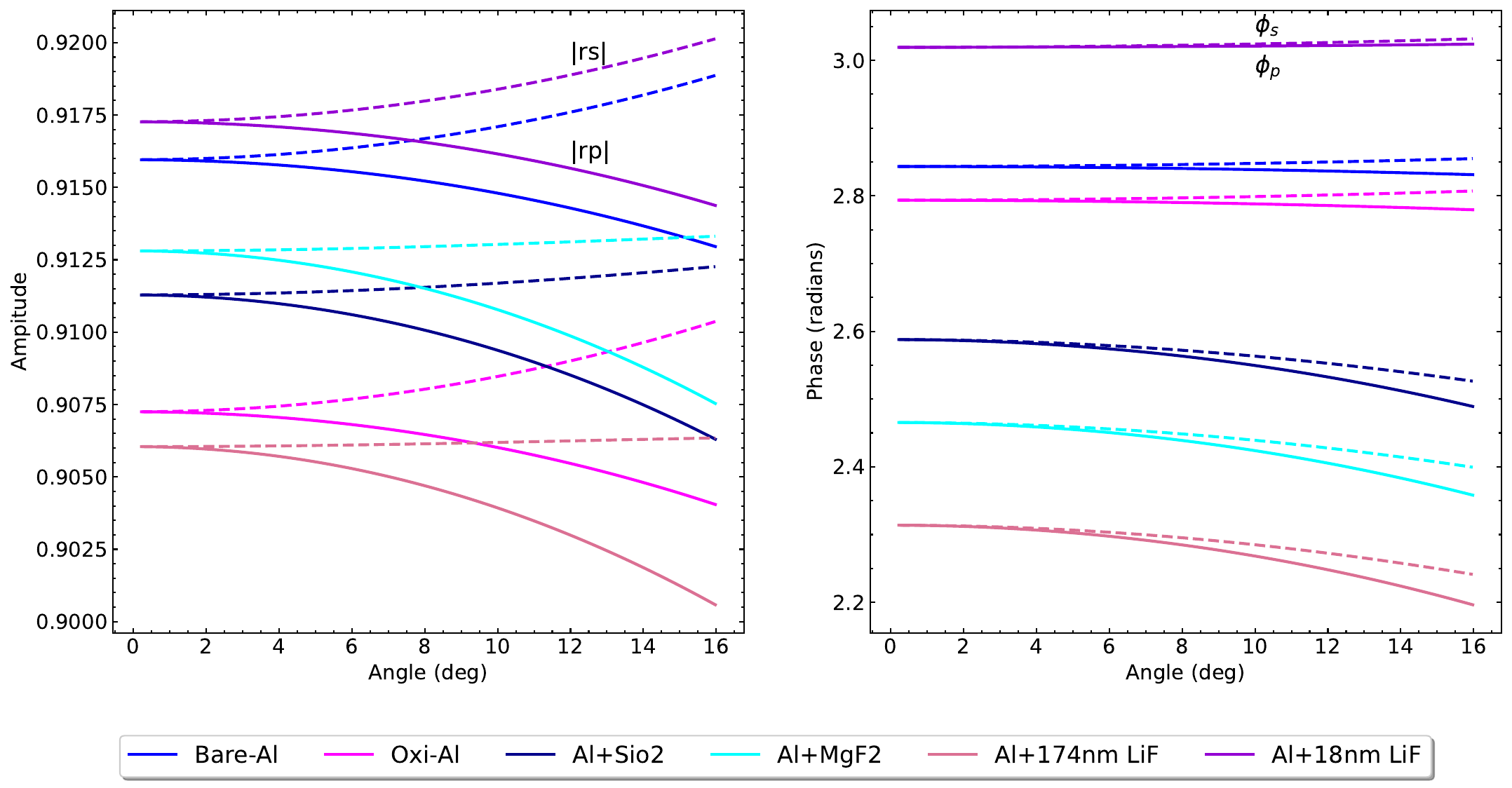}
    \caption{Amplitude and phase of the reflection coefficients for different coating recipes of aluminum and protected aluminum estimated using \textit{Macleod (2010)}\cite{macleod2010thin} algorithm. Al+ 18nm LiF coating behaves like the Bare aluminum coating.}
    \label{rprs-all-coat}
\end{figure}
Using the algorithm outlined in \textit{Macleod (2010)}\cite{macleod2010thin}, we estimate the reflection coefficients $rp$ and $rs$ for all the coatings listed in Figure \ref{ref-all-coat}. The variation of amplitude and phase $rp$ and $rs$ over the AOI is shown for a small subset of coatings that use aluminum as the main reflecting layer. Over the AOI shown, the difference in the amplitude of the reflection coefficients varies $\sim$ 0.5\% for all the coatings, and the difference in the phase is $<$ 0.5\%. The Al+ 18nm LiF coating does as well as Bare aluminum/Oxidized aluminum. It is well observed in the previous simulations in \cite{breckinridge2015polarization, anche2023polarization} that the smaller the difference in the amplitude and phase of the reflection coefficients, the smaller the magnitude of the polarization aberrations introduced in the optical system. So, Bare aluminum, Oxidized aluminum, and Al+18nm LiF are expected to have better coronagraphic performance than other coatings. 
\section{Jones pupil maps}
\label{sec2}
\begin{figure}[!h]
    \centering
    \includegraphics[width=0.95\textwidth]{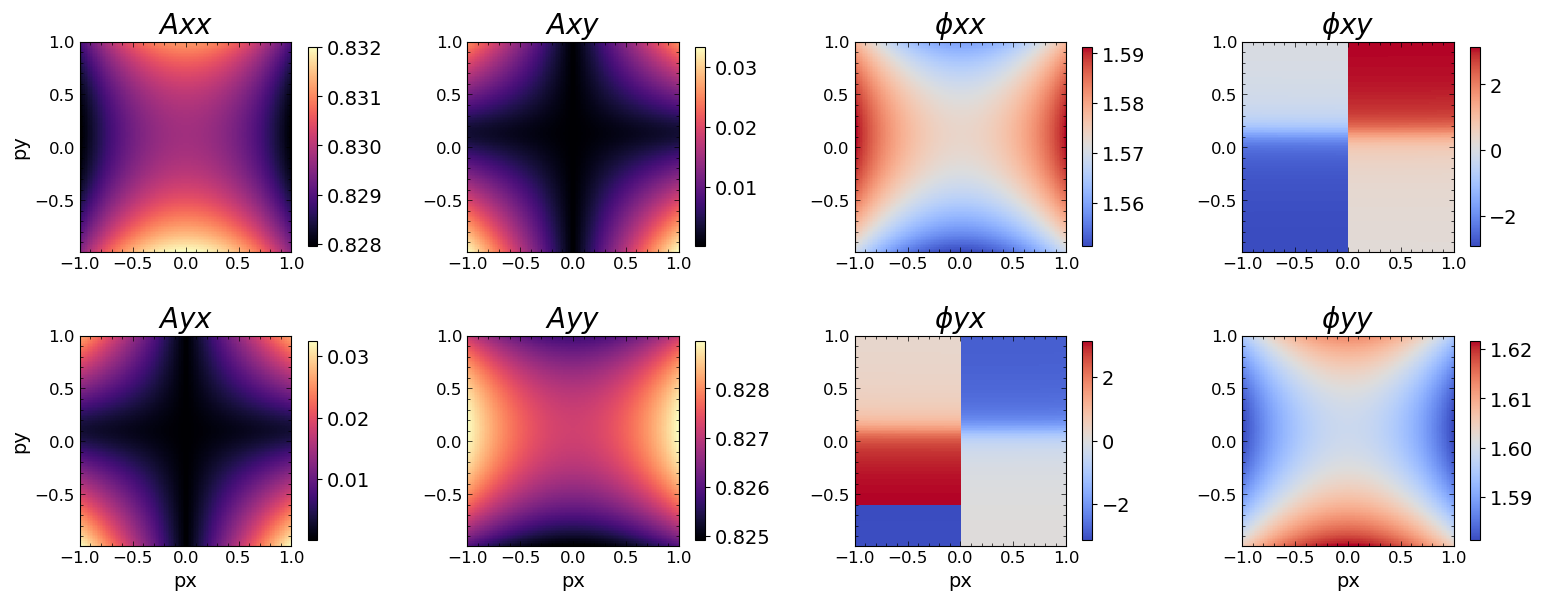}
    \caption{Jones amplitude and phase estimated at the normalized pupil coordinates shown at the exit pupil of on-axis TMA for \textit{g} band for oxidized aluminum coating. \textit{Axx, $\phi xx$ } and \textit{Axy, $\phi xy$} correspond to the output amplitude and phase for input electric field vector \textit{Ex=1, Ey=0}. Similarly, \textit{Axx, $\phi yx$ } and \textit{Axy, $\phi yy$} correspond to the output amplitude and phase for input electric field vector \textit{Ex=0, Ey=1.}}
    \label{fig:jones-stp}
\end{figure}

\begin{figure}[!h]
    \centering
    \includegraphics[width=0.95\textwidth]{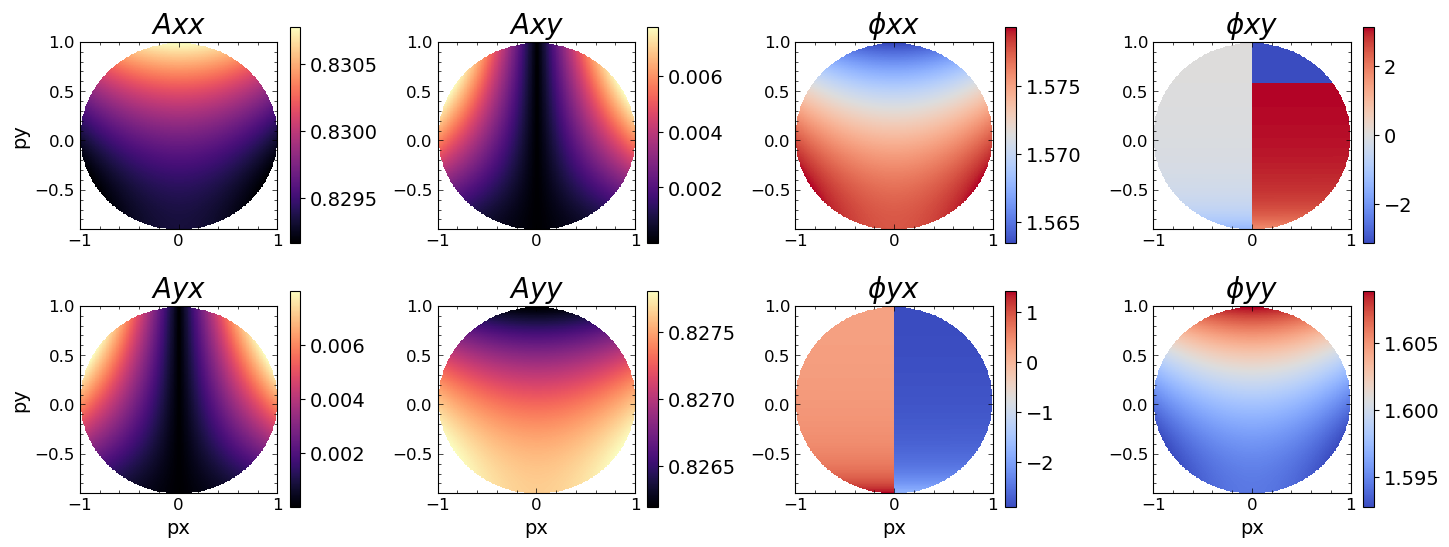}
    \caption{Jones amplitude and phase estimated at the normalized pupil coordinates shown at the exit pupil of off-axis TMA for \textit{g} band for oxidized aluminum coating. \textit{Axy} and \textit{Ayx} correspond to the cross components for inputs \textit{X-} and \textit{Y-} polarized light.}
    \label{fig:jones-tma}
\end{figure}

As a function of normalized pupil coordinates, the Jones matrices are estimated for every mirror surface, showing the evolution of polarization through the telescope. A detailed description of the Jones pupils and the PRT algorithm to estimate is provided in  \cite{chipman2018polarized}.
For on-axis TMA, we simulate the Jones pupil at the exit pupil of the telescope for eight different coatings, and for off-axis TMA, we use all the coatings shown in Figure \ref{rprs-all-coat} for astronomical \textit{g, V R, I} bands. The Jones pupil for \textit{g} band for oxidized aluminum coating is shown in Figure \ref{fig:jones-stp} for on-axis TMA and Figure \ref{fig:jones-tma} for the off-axis TMA. The amplitude \textit{Axx} and \textit{Ayy} vary over the pupil $\sim$ 0.5\% and $\sim$ 0.1\% for on-axis and off-axis TMA, respectively. The crosstalk components \textit{Axy} and \textit{Ayx} show highly apodized Maltese cross-patterns with an amplitude of $\sim$ 3.5\% and $\sim$ 0.8\% for on-axis and off-axis TMA, respectively. The phases \textit{$\phi xx$} and  \textit{$\phi yy$} show the astigmatic pattern of on-axis TMA and tilt and piston for the off-axis TMA. As seen in the case of amplitudes, the variation of \textit{$\phi xx$} and  \textit{$\phi yy$} over the pupil is larger for the on-axis TMA compared to the off-axis. These polarization-dependent phase aberrations give rise to retardance defocus, tilt, and piston in both telescopes and affect high-contrast imaging measurements, as shown in the upcoming section.
\section{Description of Coronagraph masks used and Effect on contrast}
\label{sec3}
To understand the effects of the polarization aberrations on the high-contrast imaging in these telescopes, we model three different coronagraphs: A Perfect Coronagraph (PC), a Vector Vortex Coronagraph (VVC), and a Scalar Vortex Coronagraph (SVC) using High-Contrast Imaging in Python (HCIPy) physical optics package \cite{por2018hcipy}.
\subsection{On-axis TMA}
\begin{figure}[!h]
    \centering
    \includegraphics[width=0.95\textwidth]{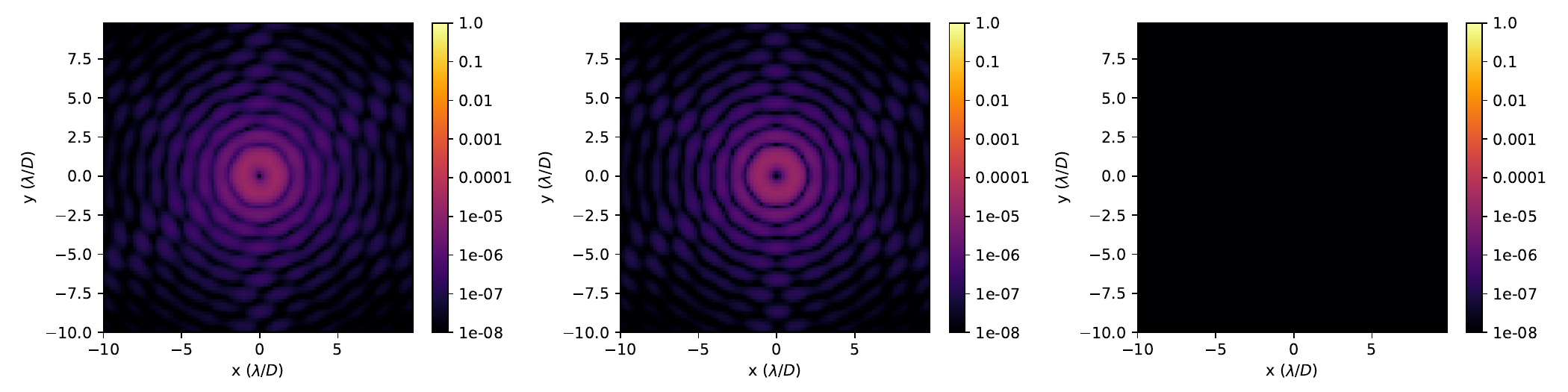}
    \caption{Stellar residuals are shown for on-axis TMA+PC coronagraph for \textit{g-} band for oxidized aluminum coating. The first, second, and third panels correspond to 2nd order coronagraph, 4th order, and 6th order coronagraph, respectively.}
    \label{fig:residuals-stp}
\end{figure}

\begin{figure}[!h]
    \centering
    \includegraphics[width=0.8\textwidth]{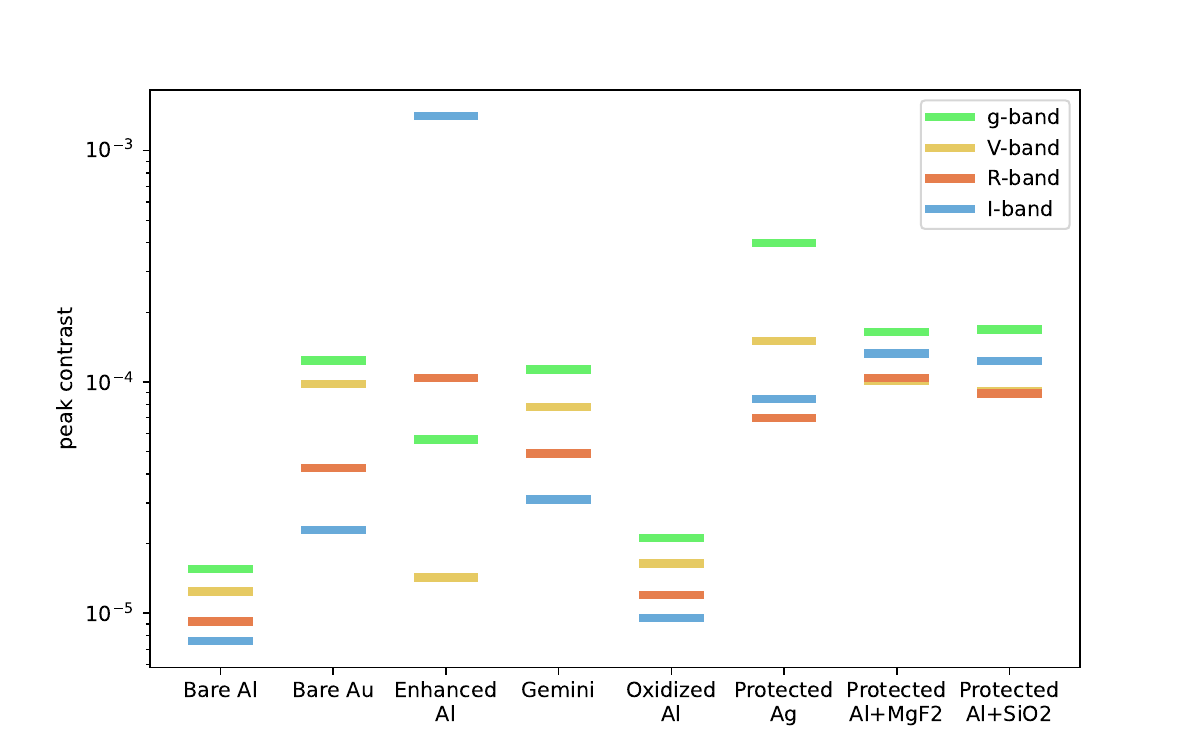}
    \caption{The peak contrast vs. different coating is shown for four astronomical filter bands for the on-axis TMA+PC model. Bare or oxidized aluminum coating performs comparatively better than the other protected coatings considered. The peak contrast correspond to the brightest speckle in the Figure \ref{fig:residuals-stp} }
    \label{fig:peakcont-stp}
\end{figure}

 The Jones pupils for the on-axis TMA were propagated through a perfect coronagraph of 2nd, 4th, and 6th order, and the resulting stellar residuals are shown in Figure \ref{fig:residuals-stp}. We estimate the peak contrast corresponding to the brightest speckle shown in Figure \ref{fig:residuals-stp} for different filter bands and coatings considered. As shown in Figure \ref{fig:peakcont-stp} for 4th order coronagraph, the peak contrast estimated for bare and oxidized aluminum coatings is in the order of $\sim 10^{-5}$, which is an order better than all the other coatings. Enhanced aluminum and protected silver perform the worst compared to other coatings. 
\subsection{Off-axis TMA}
\begin{figure}[!h]
    \centering
    \includegraphics[width=0.95\textwidth]{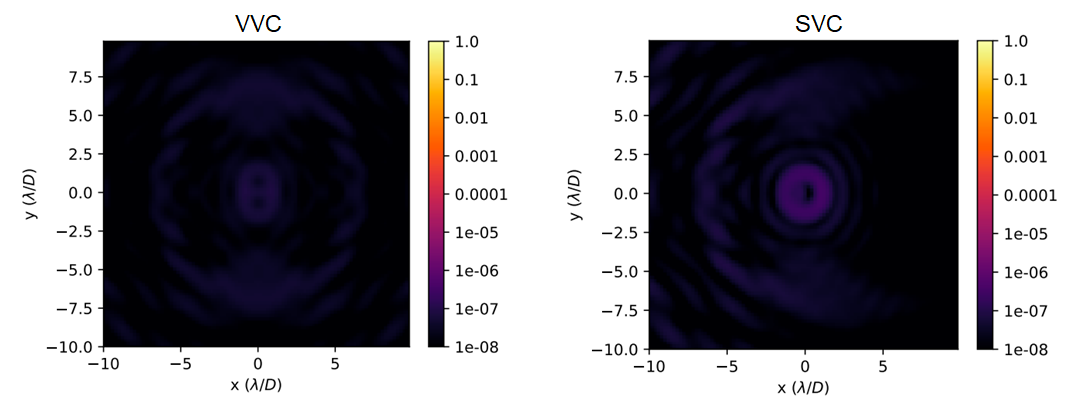}
    \caption{Stellar residuals are shown for off-axis TMA+VVC (charge-4) (left-panel) and off-axis TMA+SVC (charge-4) (right-panel) for \textit{g-} band for oxidized aluminum coating.}
    \label{fig:residuals-tma}
\end{figure}

The Jones pupils for the off-axis TMA were propagated through a VVC and SVC of charge 2, 4, and 6, and the resulting stellar residuals are shown in Figure \ref{fig:residuals-tma}. The residuals are in the order of $\sim 10^{-7}$ for VVC and $\sim 10^{-6}$ for SVC respectively. We estimate the peak contrast from the intensity of the brightest speckle for both the coronagraphs in different astronomical filter bands. 
\begin{figure}[!h]
    \centering
    \includegraphics[width=1\textwidth]{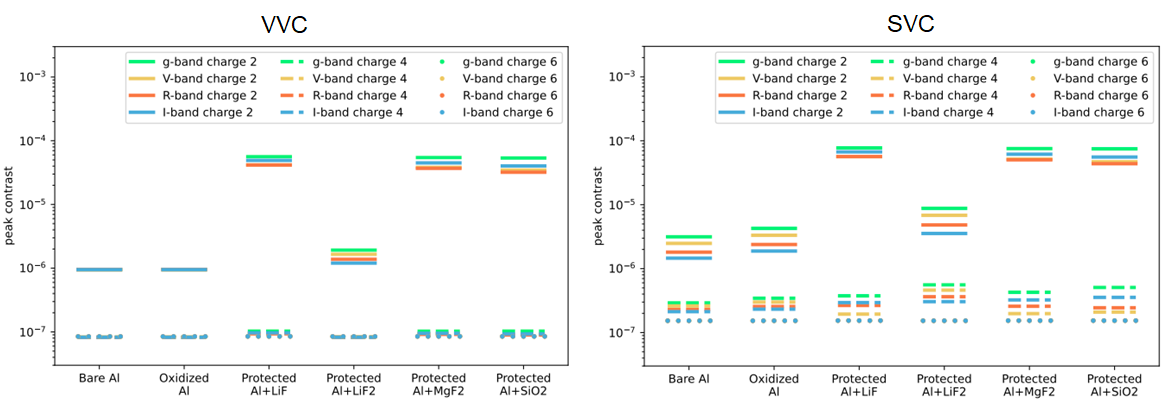}
    \caption{The peak contrast vs. different coating is shown for four astronomical filter bands for the off-axis TMA+VVC model (left panel) and off-axis TMA+SVC model (right panel)}
    \label{fig:peakcont-tma}
\end{figure}
\par
\begin{figure}[!h]
    \centering
    \includegraphics[width=1\textwidth]{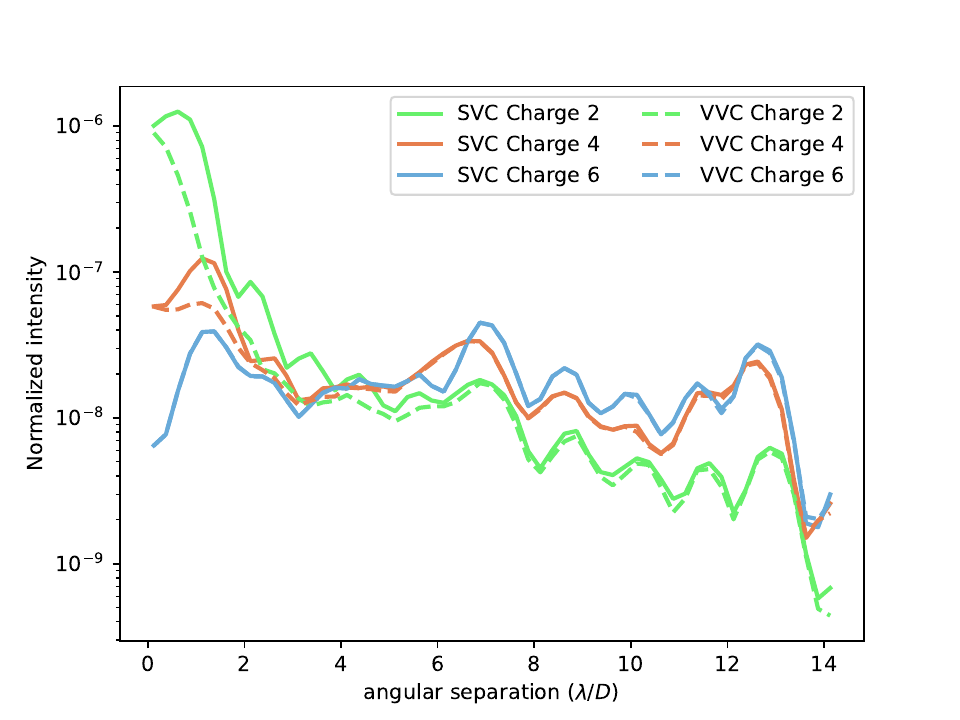}
    \caption{The radially averaged contrast vs. angular separation in \textit{g-} band is shown for the off-axis TMA (bare al)+VVC and off-axis TMA (bare al)+SVC.}
    \label{fig:radial-tma}
\end{figure}

 Bare/oxi-Al and Al+18nm LiF coating performs better than the other protected aluminum coatings. The peak contrast obtained with VVC is $\sim 10^{-7}$, and SVC is $\sim 10^{-6}$ for the bare/oxi-aluminum/Al+ 18nm LiF coating, showing that VVC performs relatively better than the SVC. Figure \ref{fig:radial-tma} compares the radially averaged contrast at different angular separations for VVC and SVC. At smaller inner working angles (IWA), VVC performs better than the SVC for charge 2 and 4 coronagraphs, but the difference closes at IWA $>$ 6$\lambda$/D.
\section{Summary}
\label{sec4}
The future space telescope Habitable World Observatory (HWO) aims to achieve a raw contrast of $10^{-10}$ to image earth-like planets. The polarization aberrations from the telescope and the instrument can introduce errors that cause the burying of the exoplanet signal.
\begin{itemize}
    \item We have modeled the polarization aberrations for two 6.5m TMA telescope designs for eight different coating recipes and propagated them into three different coronagraphs to estimate their effect on the achievable contrast.
    \item We find the dominant polarization aberrations to be retardance defocus and tilt. The crosstalk components in the Jones pupil have a higher amplitude for the on-axis TMA than the off-axis TMA.
    \item The peak contrast for the on-axis TMA with a perfect coronagraph $\sim 10^ {-5}$ for bare/oxidized aluminum is an order better than the other coatings considered. Enhanced aluminum and protected silver show worse performance than all other coatings.
    \item The peak contrast for the off-axis TMA with a VVC $\sim 10^ {-7}$ for bare/oxidized aluminum/Al+18nm LiF and off-axis TMA with a SVC $\sim 10^ {-8}$ for bare/oxidized aluminum/Al+18nm LiF. Other protected aluminum coating recipes show worse coronagraphic performance than bare/oxidized aluminum//Al+18nm LiF.
    \item For the off-axis TMA with bare/oxidized aluminum/Al+18nm LiF coating on the mirrors, Vector Vortex Coronagraph performs an order better than the Scalar Vortex Coronagraph provided, that the VVC has perfect retardance and polarization filtering.
    \item Through our analyses, we find that the Al+18nm LiF coating shows the same performance as bare/oxidized aluminum regarding reflectivity and coronagraphic performance.
    \item We will extend our analysis to incorporate a segmented primary mirror, realistic coronagraphs (VVC with polarization filtering), and wavefront control and sensing and estimate the effect of polarization aberrations on coronagraphic performance. 
\end{itemize}
 
\appendix    

\acknowledgments 
Portions of this research were supported by funding from the Technology Research Initiative Fund (TRIF) of the Arizona Board of Regents
and by generous anonymous philanthropic donations to the Steward Observatory of the College of Science at the University of Arizona  

\bibliography{ref,report} 
\bibliographystyle{spiebib} 

\end{document}